# The Isgur-Wise Limit on the Lattice[*]

Jeffrey E. Mandula[a] and Michael C. Ogilvie[b]

[a]Department of Energy, Division of High Energy Physics
Washington, DC 20585

[b]Department of Physics, Washington University
St. Louis, MO 63130

We construct the Isgur-Wise limit of QCD in a form appropriate to lattice gauge theory techniques. The formulation permits a calculation of heavy quark processes even when the momentum transfers are much larger than the inverse lattice spacing. Applications include semi-leptonic heavy quark decay and scattering processes, including the computation of the nonperturbative part of the Isgur-Wise universal function.

## 1. INTRODUCTION

The Isgur-Wise form of the heavy quark effective theory (HQET) is based on the physical idea that the motion of a very heavy object is only slightly perturbed by its interaction with much lighter objects. The idea of the Isgur-Wise limit is that a heavy quark's momentum can be usefully decomposed as $P_\mu = M v_\mu + p_\mu$, where $v$ is a constant but arbitrary four-velocity normalized to $v^2 = 1$.[1,2] It has been argued that for systems and processes that involve a single heavy quark, or one heavy quark decaying into another, the $M \to \infty$ limit gives a good approximation, and that the corrections to this limit are controlled by the ratio $\Lambda_{QCD}/M$.

The Isgur-Wise limit is based on a simple property of the fermion propagator. If we write the momentum $P$ of a heavy quark of mass $M$ as $Mv+p$, the fermion propagator simplifies in the limit $M \to \infty$ to

$$\frac{1}{\gamma \cdot P - M} \to \frac{1 + \gamma \cdot v}{2 v \cdot p} \qquad (1)$$

The power of this limit is that the large momentum $Mv$ disappears, leaving behind only terms involving $v$ and $p$, which is $O(1)$ in the limit. Another important simplification is that the Dirac matrix structure of the propagator becomes trivial. This is responsible for the spin symmetry that appears in the Isgur-Wise limit. While the same physics lies behind both the static and the Isgur-Wise forms of the HQET, the fact that $v$ can be non-zero in the Isgur-Wise form allows the calculation of processes with large momentum transfer.

The paradigm of the dynamical processes on which we are focussing is the decay of a $B$ meson into a $D$ meson plus leptons.

$$B \to D l \nu \qquad (2)$$

$$\langle D v' | \bar{c} \gamma_\mu (1 - i\gamma_5) b | B v \rangle \sim \xi(v \cdot v')$$

The Isgur-Wise universal function $\xi$ contains all non-trivial, non-perturbative information. For the vector current, the Isgur-Wise function is normalized to 1 in the forward direction (the conserved vector current). Several model calculations of the Isgur-Wise have been made, and bounds have been obtained on rather general grounds.[3,4] In addition, measurements of $B \to D l \nu$ exist from both CLEO and ARGUS, so it is quite interesting at this time to have a lattice QCD calculation of $\xi$.

Such a calculation requires the adaptation of the Isgur-Wise limit to the lattice. This entails

---

[*] Talk given at LATTICE 92, International Conference on Lattice Gauge Theory, Amsterdam (1992)

reexpressing the Isgur-Wise analysis in Euclidean coordinate space, rather than in Minkowski momentum space.[5] The key issues that must be resolved are:

a) What happens to the Minkowski space "classical velocity" $v = (v_0, \vec{v})$ in the transition to Euclidean space. The classical velocity, because of its normalization, seems to be an essentially Minkowski space concept.

b) The identification and factorization of all terms which are singular in the Isgur-Wise $M \to \infty$ limit. In Euclidean space, the propagator for a moving particle falls off more quickly than for one at rest, so one must remove rapidly decaying rather than oscillating factors.

c) As usual in the transition from Minkowski to Euclidean space, factors of $i$ can appear, and which quantities are real and which are imaginary must be identified.

We will see that in coordinate space the form that this limit takes is not that all dependence on $M$ disappears, but that it is all contained in explicit exponential factors. The practical gain is the same, though, since these can be factored out of any calculation.

The essential difference between the analysis presented here and the static treatment of heavy quarks on the lattice is the possibility of having a non-zero classical velocity.[6] That is, the heavy quark is not required to remain at a fixed site. This generalization of the static treatment would not be needed if we were to be content with computing only meson spectra or pseudoscalar decay constants.[7] However, it is needed in order to compute the Isgur-Wise function away from the forward direction, because there is no frame in which both the initial and final heavy mesons are at rest.

## 2. FREE HEAVY LATTICE FIELDS

In order to understand the form that the Isgur-Wise limit takes in a lattice context, we will illustrate it in the case of a free Dirac field with mass $M$. The mode with classical velocity $\vec{v}$ in the Isgur-Wise limit of the Euclidean-space propagator

$$S(x) = \int \frac{d^4 p}{(2\pi)^4} \frac{1}{i\gamma \cdot p + M} e^{ip \cdot x} \quad (3)$$

has the singular behavior (for $x_4 > 0$)

$$e^{iM \vec{v} \cdot \vec{x} - M v_0 x_4} \frac{1 - iv \cdot \gamma}{2} \quad (4)$$

where

$$v = (iv_0, \vec{v})$$

$$v_0 = \sqrt{1 + \vec{v}^2} \quad (5)$$

If we denote by $\tilde{S}$ the coefficient of this leading behavior, the application of the projected Dirac operator gives

$$\frac{1 - iv \cdot \gamma}{2} (\gamma \cdot \partial + M) [e^{iMv \cdot x} \frac{1 - iv \cdot \gamma}{2} \tilde{S}]$$
$$= \frac{1 - iv \cdot \gamma}{2} e^{iMv \cdot x} [-2iv \cdot \partial \tilde{S}] \quad (6)$$

The equation for the reduced propagator is simply

$$-2iv \cdot \partial \tilde{S} = \delta \quad (7)$$

All the subtleties are associated with discreteness in space, but not time, and one space dimension is sufficient to see them. The 1+1 dimensional differential-difference equation is

$$v_0 \partial_t S(t, n)$$
$$- iv_1 \frac{1}{2a} (S(t, n+1) - S(t, n-1)) \quad (8)$$
$$= \frac{1}{2a} \delta(t) \delta_{n,0}$$

where $a$ is the spacial lattice spacing. This equation is easily soluble by Fourier transforms

$$\hat{S}(t,\theta) = a \sum_n S(t,n) e^{-in\theta}$$
$$S(t,n) = \frac{1}{2\pi a} \int \hat{S}(t,\theta) e^{in\theta} \quad (9)$$

The solution is

$$\hat{S}(t,\theta) = \frac{1}{2v_0} \theta(t) e^{-\frac{v_1 t}{a v_0} \sin\theta}$$
$$S(t,n) = \frac{1}{2 a v_0} J_n\left(i \frac{v_1 t}{a v_0}\right) \quad (10)$$

It is important to note that $\hat{S}$, but not $S$ has a smooth zero spacing limit

$$\lim_{a \to 0} \hat{S}(t, p = a\theta) = \frac{1}{2v_0} \theta(t) e^{-\frac{v_1}{v_0} p t} \quad (11)$$

The exponential time dependence is exactly what is expected; it comes from the extra energy due to the additional finite momentum $\vec{p}$.

$$\lim_{M \to \infty} E(M\vec{v} + \vec{p}) - E(M\vec{v}) = \frac{\vec{v} \cdot \vec{p}}{v_0} \quad (12)$$

The fact that the Euclidean space reduced propagator has a continuum limit only in momentum space is hardly a problem, because one is always interested in hadrons with definite momentum, not the spacial position of the heavy quarks.

Similarly, the propagator of a "meson" composed of a free heavy quark and a free light particle is correctly described. The propagator of a free particle with mass $m$ in continuous time and 1 discrete space dimension is

$$\hat{s}(t,\theta) = -\frac{1}{2E} e^{-E|t|} \quad (13)$$

where

$$E = \sqrt{\frac{4}{a^2} \sin^2 \frac{\theta}{2} + m^2} \quad (14)$$

If we form a reduced composite particle propagator from the reduced heavy quark propagator and the light particle propagator

$$\Phi(t,n) = S(t,n) s(t,n) \quad (15)$$

this quantity has a smooth zero spacing limit in momentum space

$$\lim_{a \to 0} \hat{\Phi}(t, \theta = pa) = \frac{1}{2v_0} \theta(t) e^{-\frac{v_1}{v_0} p t}$$
$$\times \int_{-\infty}^{\infty} dp' e^{\frac{v_1}{v_0} p' t} \frac{e^{-t\sqrt{p'^2 + m^2}}}{\sqrt{p'^2 + m^2}} \quad (16)$$

The leading behavior of the integral is $e^{-mt/v_0}$, and the convergence of the integral is guaranteed because $v_1 < v_0$, which is intrinsic to the Euclidean space formulation of the Isgur-Wise limit.

The leading behavior of the momentum space reduced propagator is exactly what would expect of a particle with mass $(M + m)$ moving with momentum $\vec{P} = M\vec{v} + \vec{p}$, which has energy

$$E = \sqrt{(M+m)^2 + (M\vec{v} + \vec{p})^2}$$
$$\to M v_0 + \frac{m + \vec{v} \cdot \vec{p}}{v_0} \quad (17)$$

## 3. THE ISGUR-WISE FORM FACTOR

The above analysis can be immediately applied to the lattice treatment of mesons containing one heavy and one light quark, and in particular to their semileptonic decays. As above, we identify a reduced heavy quark propagator as the coefficient of the leading kinematic behavior.

$$S(x,y;A) \sim \frac{1-i\gamma\cdot v}{2} e^{iMv\cdot x} \tilde{S}(x,y;A) e^{-iMv\cdot y} \quad (18)$$

In the $M \to \infty$ limit the reduced propagator satisfies the scalar equation

$$-iv\cdot D\tilde{S} = \delta \quad (19)$$

Here $D$ is the covariant derivative. The lattice form of this derivative which automatically insures propagation in the forward time direction only is to use a forward difference in time and a symmetrical difference in space. The lattice Green's function is a solution of

$$\begin{aligned}v_0 &[ U(x,x+\hat{t})\, G(x+\hat{t},y) - G(x,y) ] \\ &+ \sum_{\mu=1}^{3} \frac{-iv_\mu}{2} [ U(x,x+\hat{\mu})\, G(x+\hat{\mu},y) \\ &\qquad - U(x,x-\hat{\mu})\, G(x-\hat{\mu},y) ] \\ &= \delta(x,y)\end{aligned} \quad (20)$$

This is soluble by forward recursion and requires scarcely more computation than the static case.

When the four-velocity is purely timelike, the operator $-iv\cdot D$ becomes $D_0$, and the static limit is recovered. In the static case the modified Dirac equation reduces to a set of decoupled one-dimensional (time) difference equations, one for each spatial point. However, when $v$ has non-vanishing space components, the equations do not decouple. Although the $\gamma$-matrix structure disappears, it involves all spatial points. In the continuum it is of course a trivium to Lorentz transform to the frame in which the heavy quark is at rest, but on the lattice no such symmetry is present, and one must deal with a finite difference equation which involves all lattice sites.

The rate for the semi-leptonic decay is contained in the three-point function $G^{(3)}(x,z,y)$ for the emission of a weak current at space-time position $x$. Removing the heavy quark propagator decay factors

$$G^{(3)} = e^{-M_1 v_1 |x_4-z_4| - M_2 v_2 |z_4-y_4|} \tilde{G}^{(3)} \quad (21)$$

gives the reduced function which is expressed in terms of lattice-calculable light and reduced heavy quark propagators. Semileptonic decay rates are computed by the standard method of taking the Euclidean time separations large and canceling the exponential decays by dividing by the appropriate two-point functions. Most of the wave function renormalizations drop out when, because of the forward normalization of the Isgur-Wise function, it is computed as the ratio

$$|\xi(v\cdot v')|^2 = \frac{G^{(3)}(v'\to v)\, G^{(3)}(v\to v')}{G^{(3)}(v\to v)\, G^{(3)}(v'\to v')} \quad (22)$$

of three-point functions for non-forward to forward decays.

By putting the Isgur-Wise limit on the lattice, not only can masses be much greater than the inverse lattice spacing, but because there is no restriction on the velocities, the momentum transfer can be large also.

**REFERENCES**

1. N. Isgur and M.B. Wise, Phys. Lett. **B232**, 113 (1989); **B237**, 527 (1990).
2. M.B. Voloshin and M.A. Shifman, Yad. Fiz. **45**, 463 (1987); Sov. J. Nucl. Phys. **47**, 511 (1988).
3. E. de Rafael and J. Taron, Phys. Lett. **B282**, 215 (1992).
4. A.V. Radyushkin, Phys. Lett. **B271**, 218 (1991).
5. J.E. Mandula and M.C. Ogilvie, Phys. Rev. D (Rap. Comm.) **45**, R2183 (1992).
6. E. Eichten and F.L. Feinberg, Phys. Rev. Lett. **43**, 1205; Phys. Rev. D **23**, 2724 (1981); E. Eichten, Nucl. Phys. B (Proc. Suppl.) **4**, 170 (1988).
7. E. Eichten, G. Hockney, and H.B. Thacker, Nucl. Phys. (Proc. Suppl.) **17**, 529 (1990).